\documentclass[aps,groupedaddress,nofootinbib,11pt]{revtex4}

\usepackage[dvips]{epsfig}
\usepackage{amsfonts}
\usepackage{amsmath}
\usepackage{}
\usepackage{color}

\definecolor{Blue}{rgb}{0.3,0.3,0.9}
\definecolor{Red}{rgb}{1.0,0.0,0.0}
\definecolor{Green}{rgb}{0,0.4,0}
\definecolor{Violet}{rgb}{0.4,0.0,0.6}
\definecolor{Cyan}{rgb}{0.0,0.4,0.6}
\definecolor{Orange}{rgb}{1.0,0.4,0.0}

\newcommand{\Eqref}[1]{(\ref{#1})}
\newcommand{\ket}[1] {\mbox{$ \vert #1 \rangle $}}

\newcommand{\bra}[1] {\mbox{$ \langle #1 \vert $}}
\newcommand{\abs}[1] {\mbox{$ \vert #1 \vert $}}

\newcounter{subequation}[equation] \makeatletter
\expandafter\let\expandafter\reset@font\csname
reset@font\endcsname
\newenvironment{subeqnarray}
  {\arraycolsep1pt
    \def\@eqnnum\stepcounter##1{\stepcounter{subequation}{\reset@font\rm
      (\theequation\alph{subequation})}}\eqnarray}
  {\endeqnarray\stepcounter{equation}}
\makeatother

\newcommand{\ba}{\begin{eqnarray}}
\newcommand{\ea}{\end{eqnarray}}
\newcommand{\sba}{\begin{subeqnarray}}
\newcommand{\sea}{\end{subeqnarray}}

\def\ch{\mbox{ch}}
\def\sh{\mbox{sh}}

\begin{document}

\vskip 1truecm

 \title{Decoherence and entropy of primordial fluctuations. \\
  I: Formalism and interpretation} 
 \vskip 1truecm
 \author{David Campo}
 \email[]{dcampo@astro.uni-wuerzburg.de}
\affiliation{Lehrstuhl f\"{u}r Astronomie,
Universit\"{a}t W\"{u}rzburg, Am Hubland,
D-97074 W\"{u}rzburg, Germany}
 \author{Renaud Parentani}
 \email[]{Renaud.Parentani@th.u-psud.fr}
 \affiliation{Laboratoire
de Physique Th\'{e}orique, CNRS UMR 8627,
Universit\'{e} Paris-Sud 11, 91405 Orsay Cedex, France}
 \begin{abstract} 
We propose an operational definition of the entropy of
cosmological perturbations based on a truncation of the 
hierarchy of Green functions. 
The value of the entropy is unambiguous despite
gauge invariance and the renormalization procedure. 
At the first level of truncation, the reduced density matrices
are Gaussian and the entropy is the only intrinsic quantity. 
In this case, the quantum-to-classical transition concerns
the entanglement of modes of opposite wave-vectors, 
and the threshold of classicality is that 
of separability.  
The relations to other criteria of classicality are established.
We explain why, during inflation, most of these criteria are not intrinsic. 
We complete our analysis by showing that all reduced density matrices  
can be written as statistical mixtures 
of minimal states, the squeezed properties of which 
are less constrained as the entropy increases.
Pointer states therefore appear not to be relevant 
to the discussion.
The entropy is calculated for various models in paper II.
 \end{abstract}
\maketitle

\section{Introduction}

The standard predictions of inflation are in noteworthy
agreement with the recent observations \cite{WMAP3}.
Yet several aspects of inflation remain poorly understood. 
Among these is the so-called 
quantum-to-classical transition of cosmological perturbations.

In the inflationary scenario,
primordial power spectra of gravitational waves and scalar perturbations 
result from the parametric amplification of vacuum fluctuations 
which begins once the modes exit the horizon.
In the course of this unitary evolution, modes 
of opposite wave-vector become more and more entangled. 
However, the primordial power spectra are impervious to this entanglement 
\cite{power,CPfat} because, for this expectation value,
the relative contribution of the quantum correlations 
is inversely proportional to the amplification factor.
Therefore, as far as (today) 
observational data are concerned, 
one can safely use a stochastic ensemble of growing modes 
in place of the pure entangled state 
predicted by the quantum treatment.

Yet, this distinction matters for other observables.
In particular, the calculation  
of the entropy and that of backreaction effects
(like any kind of radiative corrections)
must be addressed in the quantum settings. 
Although these aspects are related, in these notes we
specifically consider the question of decoherence.
We split the presentation in two papers, called I and II.
In Paper II \cite{part2}, we calculate  of the entropy in various models.
In the present paper we focus on the formulation of the question.
There is no way around this first step. 
Even though we have some notions of general features of decoherence,
there is no universal description of it. 
It occurs in a variety of ways, depending on whether the system is 
chaotic \cite{chaotic}, fermionic \cite{fermionic}, 
a two-level system \cite{twoleveldecoh}, 
or dominated by tunneling events \cite{tunnel}.
Decoherence must be analyzed case by case.

Let us therefore present the specific difficulties regarding
cosmological perturbations. 
The first is of a technical nature, the second  
concerns the consistency of the hypothesis. 
Cosmological perturbations are described by 
weakly interacting quantum fields propagating on a fixed background 
geometry.   
One therefore stumbles upon the infinities inherent to any interacting QFT.
We argue that decoherence can only be properly formulated 
in terms of expectation values of renormalized quantities. 
That is, reduced density matrices are properly  
defined by a self-consistent truncation of the hierarchy 
of Green functions,
rather than by solutions of a master equation.
The second difficulty concerns the nature of the environment.
It will be investigated in more details in \cite{part2}.
Indeed, since one is describing adiabatic perturbations,
one cannot introduce ''external'' dynamical degrees
of freedom that will act as an environment. (If one does so, 
as in \cite{Burgess,SPKnew}, decoherence obtains by construction. 
But it is based on a mechanism which might not 
be relevant in inflation, and 
it occurs at a rate which is unknown, both because the 
coupling strength and the statistical properties of the extra degrees
of freedom are {\it ad hoc}.)
The coarse graining should
be phrased in terms of the properties of the system itself.
The Green functions offer this possibility.

In these notes,
we truncate the hierarchy of Green functions at the first non trivial level: 
we retain the two-point correlation function, and  
set to zero all higher order connected correlation
functions. 
The reduced density matrices so defined are 
Gaussian and Homogeneous Density Matrices (GHDM).
These factorize in two-mode sectors characterized
by opposite wave vectors $({\bf k},-{\bf k})$.
The class of reduced density matrices being defined, 
we can turn to the calculation of the entropy and to the
analysis of the quantum-to-classical transition. 
For the entropy, we argue that its value  
is unambiguous, despite infinities of QFT and gauge invariance. 
We also prove that, during inflation, 
the entropy is the only intrinsic property of GHDM. 
Indeed, the values of the other quantities 
(e.g. the number of quanta) 
depend on the choice of the canonical variables
which are not univocally defined because
the frequency is not constant. 
We calculate the entropy for two classes of models in \cite{part2}.

To address the question of the quantum-to-classical transition,
we need to classify the reduced density matrices 
into 'classical' and 'quantum' states.
The quantum properties of GHDM are linked to the entanglement of 
modes of opposite wave vectors. 
The quantum-to-classical transition thus occurs when this
entanglement is lost 
(which happens when the state is neither pure nor thermalized,
but at a sharp frontier in between).
This gives an operational definition of 
classical states and of the time of decoherence for GHDM.

To apply this definition, one must first select creation and destruction operators.
The latter are well defined only if the Hamiltonian is time independent.
This criterion, as well as the other criteria discussed in Sec. \ref{sec:othercriteria},
are therefore unambiguous during the radiation dominated era  \cite{CPfat},
but not during inflation.
Yet, within a given representation, separability is a meaningful concept
which yields to a physical picture of entanglement.

This physical interpretation is developed in Sec. \ref{sec:othercriteria}
and \ref{sec:picture}.
In particular, in Sec. \ref{sec:picture}, 
we link decoherence to the possibility of writting GHDM 
as statistical mixtures of minimal states (displaced squeezed states),
the squeezing of which
is less and less constrained as the entropy increases. 
In Sec. \ref{sec:pointer}, we 
argue that pointer states are not 
relevant to describe 
 the decoherence of primordial fluctuations.

We have organized the paper as follows.
Sec. \ref{sec:linperturb} recapitulates the properties 
of the state of the linear perturbations.
The coarse graining is 
defined in Sec. \ref{sec:coarsegraining},
where the properties of the  
GHDM are also summarized.
Classicality is then defined as separability, 
as explained in Sec. \ref{sec:sep}.
Sections \ref{sec:coarsegraining} and \ref{sec:sep} aim also 
at clarifying our previous work \cite{CPfat}.
In Sec. \ref{sec:intrinsic}, we identify the von Neumann 
entropy as the only intrinsic statistical property of these states.
This establishes that it is unambiguous despite
the redundancy of Einstein's equations and the ambiguities 
from the perturbative renormalization of Green functions.
In Sec. \ref{sec:othercriteria}, 
the definition of separable states is compared to 
three alternative 
equivalent criteria of classicality, and to one non-equivalent criterion.
Our concluding remarks in Sec. \ref{sec:pointer}
concern the irrelevance of pointer states to 
the question of the quantum-to-classical transition.

\section{The state of linear perturbations}
\label{sec:linperturb}

\subsection{Settings}

 In models with one inflaton field, 
the dynamics of linearized (scalar and tensor) perturbations
is similar to the evolution of a massless scalar field 
$\varphi$ in the background space-time \cite{MukhaPhysRep}. 
The latter is a Friedman-Robertson-Walker space-time with flat spatial 
section. The line element is
\ba
  ds^2 = a^2(\eta) \left[ -d\eta^2 + \delta_{ij} dx^i dx^j  \right]\, ,
\ea 
The Hubble parameter $H = d\ln a/dt$ is slowly evolving. 
This variation is governed by
$\epsilon$, the first slow-roll parameter,
$ \epsilon = - d \ln H/d\ln a$. 
We consider scalar perturbations, the treatment of tensor perturbation
proceeds along similar lines. In this case, $\varphi$ 
designates the Mukhanov-Sasaki variable which is a 
linear combination of
the inflaton perturbations of the gravitational potential. 
Its Hamiltonian is unique up to a boundary term 
which corresponds to a particular choice of canonical 
variables. We will come back to this 
important point in Sec. III.C.
If we choose the conjugate momentum of $\varphi$ to be
 $ \pi = \partial_{\eta} \varphi$, 
the Hamiltonian describing the evolution of linear perturbations is 
\ba \label{Hquad}
   H &=& \int\!\!d^3 k \,  H_{\bf k,\, -\bf k} \, ,
\nonumber \\
   H_{\bf k,\, -\bf k} &=& \vert \pi_{\bf k}  \vert^2  +
\left( k^2 - \frac{\partial_{\eta}^2 z}{z}  \right) 
  \vert \varphi_{\bf k}  \vert^2 \, ,
\ea
where $\varphi_{\bf k}$ $(= \varphi^*_{-\bf k})$ is 
the Fourier transform of the field amplitude, and
$\pi_{\bf k}$ its conjugate momentum. 
The time-dependent function $z$ relates 
$\varphi$ to  $\zeta$, the scalar primordial curvature perturbation 
(defined on hypersurfaces orthogonal to the comoving worldlines \cite{MukhaPhysRep})
\ba \label{zeta}
  \varphi(t,{\bf x})= z(t) \, \zeta(t,{\bf x})
 \, , \qquad z(t)= \frac{a \sqrt{\epsilon} \, c_s}{4\pi G}\, .
\ea
In single field inflation, the sound velocity is $c_s = 1$.

We quantize each two-mode system $(\varphi_{\bf k},\varphi_{-\bf k})$
in the Schr\"{o}dinger Picture (SP)
which is best adapted to describe the Gaussian states 
considered in the following Sections. 
The field amplitude is decomposed at a given time $\eta_{\rm in}$ in terms of 
time-independent creation and annihilation operators
\sba  
  \label{phiaadagger}
  \varphi_{\bf k}(\eta_{\rm in}) &=& \frac{1}{\sqrt{2k}} \left( a^{in}_{\bf k} 
  + a^{in \, \dagger}_{-\bf k}  \right)\, , 
\qquad 
  \pi_{\bf k}(\eta_{\rm in}) = -i \sqrt{\frac{k}{2}} \left( a^{in}_{\bf k} 
  - a^{in \, \dagger}_{-\bf k}  \right)\, ,
\sea
where $a^{in}$ and $a^{in \, \dagger}$ verify the commutation relations
\ba
  \left[a^{in}_{\bf k} ,\,  a^{in \, \dagger}_{\bf k'} \right] = 
  \delta^{(3)}\left( {\bf k} -  {\bf k}' \right)\, .
\ea
The free vacuum will be taken to be the Bunch-Davis (BD) vacuum, 
defined as the state which minimizes the Hamiltonian \Eqref{Hquad} 
in the asymptotic past
\ba \label{BDSP}
  &&a^{in}_{\bf k} \ket{0 \, {\rm in}} = 0 \, , \qquad  {\rm for} \quad 
  \eta_{\rm in} \to -\infty \, .
\ea
Alternately, 
in the Heisenberg Picture (HP), it corresponds to the state with 
only positive frequencies in the asymptotic past.
In the decomposition of the field amplitude 
$\varphi_{\bf k}(\eta) = a^{in}_{\bf k} \varphi_k^{in}(\eta) 
  + a^{in \, \dagger}_{-\bf k} \varphi_k^{in\, *}(\eta)$,
the mode functions $\varphi_k^{in}$ verify
\ba \label{BDHP}
  \left( i \partial_{\eta} - k  \right)\varphi_k^{in} \vert_{k\eta \to -\infty}
   = 0\, .
\ea
In a de Sitter background, $a = -1/H\eta$, they are given by
\ba \label{modedS}
  \varphi_k^{in}(\eta) = 
   \frac{1}{\sqrt{2k}} \left(1 - \frac{i}{k\eta} \right) e^{-ik\eta} \, .
\ea
As is well known, there are no linear scalar perturbations 
in a purely de Sitter background. The solution \Eqref{modedS} merely
serves the purpose to write  
the modes in a closed form. In the slow-roll approximation,
the scalar perturbation spectra in the long wavelength limit ($k\eta \ll 1$) 
can be inferred from the above solutions by the substitution
$H \mapsto H_k/\sqrt{\epsilon_k}$ where the background quantities 
are evaluated at horizon crossing, i.e. $k=a_kH_k$.

The free evolution corresponding to the Hamiltonian (\ref{Hquad}) 
preserves the Gaussianity and the purity of  
the initial state \Eqref{BDSP}.
In addition, each two mode sector 
can be analyzed independently since the state and the Hamiltonian
split into a tensor product and a sum respectively,
\ba \label{factorization}
  \ket{\Psi_{\rm in}(\eta)} &=& \bigotimes_{\frac{1}{2} {\bf k}'s} 
  \ket{\Psi_{{\bf k}, -{\bf k}}(\eta)}\, ,
\nonumber \\
  \ket{\Psi_{{\bf k}, -{\bf k}}(\eta)} &=& {\cal T} 
  \exp\left( -\int_{-\infty}^{\eta}\!\!d\eta' \, H_{\bf k,\, -\bf k}(\eta') \right)
 \ket{0\, {\rm in},({\bf k}, -{\bf k})}\, ,
\ea
where the tensor product is over half of the wave vectors 
since $\ket{\Psi_{{\bf k}, -{\bf k}}}$ is a two-mode state.

\subsection{The covariance matrix}

Because the pairs of modes are statistically independent
in the state \Eqref{factorization}, we consider only
one such pair and drop the index $({\bf k}, -{\bf k})$.
To use only one formalism throughout the paper, from now on we adopt 
 the density matrix notation 
\ba 
  \rho(\eta) = \ket{\Psi} \bra{\Psi}\, .
\ea

Since $\rho(\eta)$ is Gaussian, its statistical properties
are summarized in the 
one and two-times correlation functions.
The former vanish identically by statistical homogeneity.
Many of the second moments vanish as well, 
also because of statistical homogeneity, namely
${\rm Tr}\left( \rho \varphi_{\pm \bf k}^2  \right) = 
{\rm Tr}\left( \rho \pi_{\pm \bf k}^2  \right) = 
{\rm Tr}\left( \rho \varphi_{\bf k} \pi_{\bf k}  \right) = 
{\rm Tr}\left( \rho \varphi_{-\bf k} \pi_{-\bf k}  \right) = 0$.
Finally, due to the relation 
$\varphi_{\bf k}^{\dagger} = \varphi_{-{\bf k}}$ 
and similarly for $\pi_{\bf k}$,
the remaning covariances can be conveniently condensed into a  
$2\times2$ matrix (instead of $4\times4$) that we shall 
also call the covariance matrix
\ba \label{covmatrix}
    C &\equiv& \frac{1}{2}
  {\rm Tr} \left(  \rho \, \left\{ V ,\, V^{\dagger} 
   \right\} \right)
  =    
\left(
   \begin{array}{cccc}
          {\cal P}_{\varphi}  &  {\cal P}_{\varphi \pi} \\
            {\cal P}_{\varphi \pi} &  {\cal P}_{\pi}   
   \end{array} 
   \right) \,, 
 \qquad 
  V =    
\left(
   \begin{array}{c}
          \sqrt{k} 
          \varphi_{\bf k}\\
          \pi_{-{\bf k}}/\sqrt{k}        
   \end{array} 
   \right) \, ,
\ea
where the ${\cal P}$'s are functions of $\eta$ and $k$. 
In the inflationary phase (approximated by a de Sitter evolution), 
they are given by
\sba \label{covmatInf}
  {\cal P}_{\varphi} &=&  \frac{k}{2} 
  \langle \left\{ \varphi_{\bf k},\, \varphi_{-{\bf k}} \right\} \rangle 
  = k \abs{\varphi_{k}^{in}}^2
  = \frac{1}{2}\left( 1 + \frac{1}{x^2}  \right) \, ,
  \\
   {\cal P}_{\pi} &=&  \frac{1}{2k} 
   \langle  \left\{ \pi_{\bf k},\, \pi_{-{\bf k}} \right\} \rangle  
   = k^{-1} \abs{\partial_{\eta}\varphi_{k}^{in}}^2
  = \frac{1}{2}\left( 1 - \frac{1}{x^2} + \frac{1}{x^4}  \right)\, ,
 \\
   {\cal P}_{\varphi \pi} &=&  \frac{1}{2} 
   \langle \left\{ \varphi_{\bf k},\, \pi_{-{\bf k}} \right\} \rangle  
   = 2 Re\left[\varphi_{k}^{in\, *}\, \partial_{\eta}\varphi_{k}^{in}\right]
  = -\frac{1}{2x^3}\, ,
\sea
where $\{\, ,\, \}$ is the anticommutator 
and we used the notation $x = k\eta$.

We introduce an additional 
representation of the state \Eqref{factorization} which clearly displays
the entanglement between modes of opposite wave vectors: 
\sba
\label{defn}
  n &\equiv& {\rm Tr}\left( \rho(\eta) \, 
  a_{\bf k}^{{\rm in} \, \dagger} a_{\bf k}^{\rm in}   \right) \, , 
\\
\label{defc}
  c &\equiv& 
    {\rm Tr}\left( \rho(\eta) \, a_{\bf k}^{\rm in} a_{-\bf k}^{\rm in}   \right) \, , 
\sea
$n$ is real while $c$ is complex.
The utility of this represention stems from the fact that $n$ is simply 
related to the power spectrum while
$\vert c \vert$ measures the strength of the correlations between 
the two modes.
These three real numbers provide
an equivalent representation of the covariance matrix since
inverting Eqs. \Eqref{phiaadagger} and 
inserting into \Eqref{defn} gives
\ba
\label{n+1/2}
  n + \frac{1}{2} &=& \frac{1}{2} \left( {\cal P}_{\varphi}+{\cal P}_{\pi} \right) 
 \, , \quad 
 {\rm Re}(c) = \frac{1}{2} \left( {\cal P}_{\varphi} - {\cal P}_{\pi}\right) 
 \, , \quad 
  {\rm Im}(c) = {\cal P}_{\varphi \pi}\, .
\ea

As we shall see in Sec. \ref{sec:propGHDM}, the determinant of $C$ is related
to $S$, the von Neumann entropy of the state.  
With \Eqref{covmatInf}, one finds that the $x$ (time) dependence
drops out and get 
\ba
  \det(C) = \frac{1}{4}  \quad \Longleftrightarrow \quad 
  \vert c \vert^2 = n(n+1) \quad \Longleftrightarrow \quad 
  S = 0\, .
\ea 
 The minimal value that $\det(C)$ can take is ${1}/{4}$.
When this lower bound is saturated,  
the state minimizes the Heisenberg uncertainty relations since
$\det(C) = {\cal P}_{\varphi} {\cal P}_{\pi} - {\cal P}_{\varphi \pi}^2 \geq 1/4$.

The state of the linear perturbations is therefore characterized 
by a single function,
the power spectrum ${\cal P}_{\varphi}(k,\eta)$ 
(and an angle that plays no part in this paper).
In realistic models, i.e. with interactions, the complete description of the system
requires the full hierarchy of connected correlations functions.
Such knowledge is out-of-reach, and in practice
one resorts to a coarse grained description. 
This gives a reduced state $\rho_{\rm red}$ characterized by a non zero entropy.
The choice of the correlations that are discarded must be done 
on physical grounds and in a way consistent with the dynamics. 
This is the subject of the next Section.

\section{Coarse graining and reduced Gaussian density matrices} 
\label{sec:coarsegraining}

In \ref{sec:opdef}, we present an operational definition of 
reduced density matrices 
appropriate for interacting field theories.
The advantages of this definition will be emphasized in paper II
through the analysis of explicit models.
We apply it to cosmological perturbations in \ref{sec:applycosmo}
in the simplest case, the Gaussian approximation.
The properties of the Gaussian states are summarized in \ref{sec:propGHDM}.

\subsection{Operational definition of coarse graining}
\label{sec:opdef}

The program begins with the specification of 
a finite set of ''relevant'' observables 
$\left\{ \hat {\cal O}_{1}, ..., \hat {\cal O}_{n}  \right\}$
(which can be functions of both time and space or momentum). 
This set defines our knowledge about the state
and the dynamics (see Appendix \ref{sec:whybother}
where this important aspect is emphasized) of the system. 
In general the finite set 
$\left\{ \hat {\cal O}_{1}, ..., \hat {\cal O}_{n}  \right\}$
allows only for a partial reconstruction of this state.  
The reconstruction is performed in the following way.
The {\it reduced} density matrix 
$\rho_{\rm red}\left( \bar {\cal O}_{1}, ...,  \bar {\cal O}_{n} \right)$ 
is defined by the constraints on
the expectation values $\bar {\cal O}_{j}$ of these observables,
\ba \label{constraints}
  &&{\rm Tr}\left( \rho_{\rm red}  \right)  = 1 \, , \nonumber \\
  &&{\rm Tr}\left[\rho_{\rm red}
  \left( \bar {\cal O}_{1}, ...,  \bar {\cal O}_{n}
  \right)  \hat {\cal O}_{j} \right] 
  \equiv \bar {\cal O}_{j} \, .
\ea
There is of course an infinite number of density matrices verifying these constrains.
However, to be consistent with our hypothesis that the $\hat {\cal O}_j$ 
are the only 
observables accessible to us, we must choose the density matrix 
$\rho_{\rm red}$ 
which maximizes the 
entropy given the constraints \Eqref{constraints}:
 \ba \label{defentropy}
   S\left[ \rho_{\rm red} \right] \geq S\left[ \rho \right]\, ,
 \ea
where 
\ba
  S\left[ \rho \right] = -{\rm Tr}\left(  \rho \ln \rho \right)\, .
\ea
The formal solution of this variational problem is \cite{Robertson} 
\ba \label{rhoformal}
  \rho_{\rm red} = \frac{1}{Z_{\cal O}} \exp\left( -\sum_{j=1}^n 
   \lambda_j  \hat {\cal O}_n \right)
   \, , \qquad 
   Z_{\cal O} = {\rm Tr}\left[ \exp\left( -\sum_{j=1}^n \lambda_n  
  \hat {\cal O}_j \right)  \right]\, .
\ea
(In the case these observables depend on space-time, one should read
$\int\!\!d^3x \, \lambda(t,{\bf x})\hat {\cal O}_n(t,{\bf x})$).
This is an out-of-equilibrium generalization of Gibb's canonical and
grand canonical states which reduce to  
these distributions in equilibrium.

The $\lambda_n$ are Lagrange multipliers.
The constraints on the expectation values \Eqref{constraints} are therefore 
written
\ba \label{varZ}
  \bar {\cal O}_{j} = - \frac{\partial}{\partial \lambda_j} \ln Z_{\cal O}\, .
\ea
One then inverts this system of $n$ equations to 
express the Lagrange multipliers in terms of
the expectation values
\ba
  \lambda_j = \lambda_j\left( \bar {\cal O}_{1}, ..., \bar {\cal O}_{n}  \right)\, .
\ea
and substitute in \Eqref{rhoformal}.
The von Neumann entropy of 
the solution $\rho_{\rm red}$ is the Legendre transform 
of the logarithm of the partition function $Z_{\cal O}$,
\ba \label{entropyformal}
  S[\rho_{\rm red}] = \ln Z_{\cal O} + 
  \sum_{j=1}^n \lambda_j  \bar {\cal O}_j = 
  S\left(  \bar {\cal O}_{1}, ..., \bar {\cal O}_{n}  \right)\, .
\ea
For quantum fields, one notices the close resemblance between $\ln Z_{\cal O}$
and the generating functional of Green functions, and between
$S$ and the effective action.

Notice that the solution \Eqref{rhoformal} is formal. Ambiguities
stem from the non-commutativity of the operators $\hat O_j$. 
As a result, \Eqref{entropyformal} is strictly 
valid only when all the $\hat O_j$ commute (see also 
Eq. \Eqref{stupidentropy} and the following comment).
In the following we will only consider Gaussian density matrices. 
The theory of their representation is well developed and we shall
rely on this body of work. In particular, 
the ordering ambiguities in \Eqref{rhoformal} and \Eqref{entropyformal}
can be resolved. The correct formula of the entropy is 
given by Eqs. \Eqref{SvN} and \Eqref{SvN2}.

\subsection{Application to quantum fields}

For quantum fields, the above program transposes into the following.
We start from the observation that 
the knowledge of the Green functions of a (self-)interacting field 
is equivalent to the knowledge of the state of that field.
A coarse graining is therefore naturally defined
 by a truncation of the (BBGKY) 
hierarchy of Green functions at a given rank $N$. 
As explained above, to be self-consistent, 
the hierarchy of Green functions must be closed.

This coarse graining is the field theoretic version of 
Boltzmann's {\it Ansatz}. In the latter, $N$-body interactions are
neglected for $N\geq 3$ and the object of physical interest is the
one-particle correlation function.  
Beyond the Gaussian approximation, 
this coarse graining is formalized in terms of the so-called
$n$-Particle Irreducible 
representations of the effective action \cite{HuBBGKY}.

\subsection{Coarse grained description of metric perturbations}
\label{sec:applycosmo}

Let us apply this program
to the scalar perturbation $\varphi$. 
The lowest non trivial order of truncation is at $N=2$.
The reduced density matrix is then defined by the anticommutator of 
$\varphi$.
The corresponding reduced density matrices $\rho_{\rm red}$
which maximize $S$ are Gaussian \cite{maxGaussian}. 
These states will be described in details in the next Sections.
Here we wish to give a qualitative understanding of this coarse graining.

Each Fourier 
component of the anticommutator describes the effective evolution
of the two mode sector 
$(\varphi_{\bf k},\varphi_{\bf -k})$
and can be analyzed separately, as it was the case in the 
linearized treatment.
The growth of entropy associated with the coarse graining
in the present case can be therefore described
by  the set of Gaussian  
two-mode density matrices $\rho^{\rm red}_{{\bf k}, {\bf -k}}$.  
Each of them characterizes
 the loss of entanglement between the 
two modes in the presence of interactions, which
are self-interactions of gravity or (and) 
interactions of the metric perturbations with other fields 
(e.g. the fields of the Standard Model of particle physics or  
isocurvature perturbations in multi-inflaton field scenarios).
The environmental degrees of freedom
are thus either the collection of the modes 
$\varphi_{{\bf k}' \neq {\bf k}}$, or modes of other fields.

Returning to the formalism, for each two-mode,
$\rho^{\rm red}_{{\bf k}, {\bf -k}}$ 
is defined by the anticommutator 
\ba \label{defrhored}
  G_a(\eta,\eta';k) &=&  \frac{1}{2}
  \int\!\!d^3x \, e^{i{\bf k}{\bf x}}  \, 
  {\rm Tr} \Big[ \rho_{\rm tot}(\eta_{in})\,   
  \left\{ \varphi(\eta,{\bf x}) \,,  
   \varphi(\eta',{\bf 0}) \right\} 
    \Big]\, ,
\ea   
where the in-state 
$\rho_{\rm tot}(\eta_{in})$ is the interacting vacuum of the total system.
It replaces the Bunch-Davis vacuum of linearized modes. 
$\varphi(\eta,{\bf x})$ is the 
Heisenberg operator of the nonlinear metric perturbations. 
In the SP, using (\ref{covmatrix}), 
$\rho_{\rm red}(\eta)$ is the Gaussian density matrix
possessing the following covariance matrix
\ba \label{covmatdressed}
  {\cal P}_{\varphi} &=& G_a(\eta,\eta;k) 
  \, , \quad 
  {\cal P}_{\varphi \pi} = \partial_{\eta'}G_a(\eta,\eta';k)\vert_{\eta=\eta'} 
  \, , \quad  
  {\cal P}_{\pi} = 
  \partial_{\eta}\partial_{\eta'}G_a(\eta,\eta';k) \vert_{\eta=\eta'}  \, . 
\ea
It will indeed be shown in \cite{part2} that in the Gaussian approximation
it is always possible to make a canonical transformation
$(\varphi,\pi') \mapsto (\varphi,\pi = \partial_{\eta} \varphi)$. 
As recalled in Sec. \ref{sec:intrinsic}, the entropy is invariant under canonical
transformations. 
Throughout this paper we use this Gaussian approximation. 
Since the interactions do not spoil the property of statistical homogeneity, 
the reduced states can be described in the same way as the pure state of the 
linear perturbations [Sec. \ref{sec:linperturb}]. 
Only the values of the covariance matrix element differ. 
These are now given by 
\Eqref{covmatdressed} in place of \Eqref{covmatInf}.
We call these states Gaussian Homogeneous Density Matrices (GHDM).

\subsection{Entropy of $\rho_{\rm red}$}
\label{sec:propGHDM}

To measure the strength of the correlations 
between ${\bf k}$ and $-{\bf k}$,
we introduce the parameter $\delta$ defined by
\ba
  \vert c \vert^2 \equiv n(n+1 - \delta)\, ,\qquad  0 \leq \delta \leq n+1 \, .
  \label{ldef}
\ea
The standard inflationary distribution of Section II
is maximally coherent and corresponds to $\delta = 0$.
The least coherent distribution 
corresponds to $\delta = n+1$.
The entropy \Eqref{defentropy} is a strictly growing function of $\delta$.
It is related to the determinant of the covariance matrix by
\ba \label{SvN}
  S &=&  - {\rm Tr}\left( \rho \ln \rho  \right) 
  = 2\left[(\bar n + 1 ) \ln(\bar n + 1) - \bar n \ln(\bar n)\right] \, ,
\ea
where the parameter $\bar n$ is defined by
\ba \label{SvN2}
   \left( \bar n + \frac{1}{2} \right)^2 
   \equiv  {\rm det}(C)  
   = \frac{1}{4} + n \delta 
 \, . 
\ea 
The prefactor $2$ in \Eqref{SvN}
accounts for the fact that $\rho$ is the state of two modes.
In the range 
$\delta \gg 1/n$, $n \gg 1$, the expression \Eqref{SvN} of the entropy
simplifies,
\ba \label{Sdelta}
  \bar n \simeq \sqrt{n \delta} \, , \qquad 
  S = \ln\left( n \delta \right) + O(1) \, .
\ea
These equations have a simple geometric interpretation that we will use 
in Sec. \ref{sec:othercriteria} and \ref{sec:picture}.
The instantaneous eigenvectors of the covariance matrix are given by 
the rotated canonical variables
 \ba \label{rotatedvar}
   \Phi_{\bf k}(\eta) &=& \cos\theta\, \sqrt{k}\varphi^{in}_{\bf k} 
   + 
   \sin \theta\,
   \, \frac{\pi^{in}_{\bf k}}{\sqrt{k}} 
   = \frac{1}{\sqrt{2 }} (\hat a_{\bf k}^{in} e^{-i\theta} +
   \hat a_{-\bf k}^{in\, \dagger}e^{+i\theta}) \, ,
   \nonumber \\
   \Pi_{\bf k}(\eta) &=&   - \sin\theta \,\sqrt{k} \varphi^{in}_{\bf k}
   + \cos\theta
   \,  \frac{\pi^{in}_{\bf k}}{\sqrt{k}} 
   = \frac{-i}{\sqrt{2}} (\hat a_{\bf k}^{in} e^{-i\theta} -
   \hat a_{-\bf k}^{in\, \dagger}e^{+i\theta})\, .
 \ea
The angle $\theta = \theta(\eta)= \frac{1}{2} \arg(c(\eta))$ 
gives the orientation of the eigenbasis of the covariance matrix 
with respect to the (fixed) 
original variables $\varphi^{in}_{\bf k}$ and 
$\pi^{in}_{- \bf k}$.
The eigenvalues of the covariance matrix are 
the variances of $\Phi$ and $\Pi$, 
\sba \label{eigenvalues}
  \langle \{ \Phi_{\bf k} , \, \Phi^\dagger_{\bf k} \} \rangle &=& 
   n+\frac{1}{2}  + \abs{c} = 2 n + O(1, \delta) \, , 
\\
\label{eigenPi}
   \langle \{ \Pi_{\bf k} , \, \Pi^\dagger_{\bf k} \} \rangle &=&
   n+\frac{1}{2} - \abs{c} = \frac{\delta}{2} + O\left(  \frac{1}{n} \right) \, ,
\sea
and $\langle \{ \Phi_{\bf k} , \, \Pi^\dagger_{\bf k} \} \rangle = 0$.
Hence, when $\delta \gg 1/n$, one has
\ba
  \det(C) \simeq n\delta \simeq {\cal A}\quad \Longrightarrow \quad 
 S \simeq \ln{\cal A} \, ,
\ea
where ${\cal A}$ is the area under the 
$1\!\!-\!\!\sigma$ contour in phase space.

\section{The intrinsic properties of GHDM}
\label{sec:intrinsic}

The von Neumann entropy $S = - {\rm Tr}\left(\rho \ln(\rho)  \right)$ 
is manifestely an intrinsic property of the state.
We show that for GHMD it is the only intrinsic property.
Important consequences are then derived.

\subsection{Entropy as the unique intrinsic property}

To illustrate the question, let us first consider the following situation.
In place of the Mukhanov-Sasaki variable, one could choose instead to 
work directly with the  
curvature perturbation $\zeta$. The quadratic part of the 
Lagrangian is
\ba \label{Squad}
    S_\zeta &=& 
  \frac{1}{8\pi G}\int\!\!dt d^3x \, \epsilon(t) \, 
 [a^3(\dot \zeta)^2 - a (\nabla \zeta)^2]\, ,
\ea 
where $\dot \zeta = \partial_t \zeta = a^{-1} \partial_{\eta} \zeta$.
The conjugate momentum of the curvature perturbation is 
\ba \label{pisol}
  \pi_{\zeta}(t) = \frac{a^3 \epsilon}{4\pi G} \, \dot \zeta(t) \, .
\ea
Assuming the slow roll condition $\epsilon \simeq {\rm cte}$
and $\dot \epsilon/H\epsilon \simeq {\rm cte}$, 
the modes with positive frequency in the asymptotic past are
\ba \label{modesol}
  \zeta_k(\eta) = \zeta_k^0 \left( 1+ i k\eta \right) e^{-ik\eta} \, , \qquad 
  \vert \zeta_k^0 \vert^2 = \frac{4\pi G }{\epsilon_k} \, 
  \frac{H_k^2}{2k^3}\, .
\ea
From this solution and \Eqref{pisol}, one obtains the following expressions for
the covariance matrix at tree level
\sba 
\label{Pzeta}
  {\cal P}_{\zeta} &\equiv& \frac{1}{2}
  \langle  \{ \zeta_{\bf k}(\tau), \, \zeta_{-{\bf k}}(\tau) \}\rangle
 = \abs{\zeta_k^0}^2 \left( 1+ x^2 \right) \, ,
\\ 
\label{Pzetapi}
  {\cal P}_{\zeta \pi} &\equiv& \frac{1}{2}
   \langle \{ \zeta_{\bf k}(\tau), \, \pi_{-{\bf k}}(\tau)\} 
   \rangle 
 = - \frac{a^3 \epsilon}{4\pi G} \,H x^2 \abs{\zeta_k^0}^2  \, ,
   \\
\label{Ppi}
   {\cal P}_{\pi} &\equiv&  \frac{1}{2} 
   \langle \{ \pi_{\bf k}(\tau),\, \pi_{-{\bf k}}(\tau) \}\rangle
 =  \left( \frac{a^3 \epsilon}{4\pi G}
 \right)^2 H^2 x^4 \abs{\zeta_k^0}^2  \, .
\sea
Despite the differences with the variances \Eqref{covmatInf}, one checks that
$ \det(C) = {1}/{4}$. 
The determinant of the covariance matrix, hence the entropy, are invariant 
under the linear canonical transformation 
$(\varphi,\pi) \mapsto (\zeta,\pi_{\zeta})$.

More generally, the entropy 
does not depend on the choice of canonical variables.
Indeed, canonical transformations are linear symplectic 
transformations of the covariance matrix.
The intrinsic properties of $\rho_{\rm red}$ 
are therefore the symplectic invariants (of the group 
${\rm Sp}(4, \textbf{R})$ for a system of two modes). 
The symplectic invariants of a $4\times 4$ covariance matrices
are known \cite{sympinv}.
When the constraint of statistical homogeneity is added, 
these invariants are degenerate into a single quantity, namely 
the determinant of $C$. In other words, $\det(C)$ is the unique 
intrinsic property of the effective Gaussian state $\rho_{\rm red}$ 
of the cosmological perturbations.
Unique here is employed in the sense of an equivalence class: 
any quantity which can be expressed in terms of $\det(C)$ 
only, e.g. the entropy,
characterizes the same quantity.
It is interesting to notice that this uniqueness rests on  
statistical homogeneity.

We now apply this result to prove that the entropy is well defined 
despite the redundancy of Einstein's equations and the arbitrariness in the
definition of the perturbatively renormalized Green functions.

\subsection{Entropy and gauge invariance}

The entropy of metric perturbations is independent of 
the choice of gauge. The reason is that in a change of coordinates 
$x^{\mu} \mapsto x^{\mu} + \xi^{\mu}$, 
the Lagrangian (of Einstein-Hilbert plus inflaton) 
changes by a total derivative.
 It can therefore be seen as a canonical transformation.
That the entropy is a gauge invariant quantity can also be seen independently
from the fact that we quantize gauge invariant variables like 
the Mukhanov-Sasaki variable or the comoving curvature perturbations.

\subsection{Entropy and renormalization}

To show the independence of the entropy of the renormalization scheme, we 
proceed in three steps.
We consider first renormalizable field theories 
in Minkowski space, then generalize to non 
renormalizable theories in Minkowski space,
and finally to non renormalizable theories on a curved background.
In what follows we are only considering the entropy per two-mode.

For renormalizable theories in Minkowski space, we have to 
examine two 
potential sources of ambiguities, the parameters of the 
Lagrangian and the wave function renormalization.
First consider the parameters. They are local terms in the effective action,
and therefore appear as local terms in the equation of the propagator.
It can be checked (see paper II) 
that they are not responsible for the change of entropy.

In a renormalizable theory, after renormalization of the bare parameters, the 
renormalized Green's functions are numerically equal to the bare ones, up 
to a multiplication factor given by the wave function renormalization constant(s)
to the appropriate power.
The field strength renormalization is also a canonical transformation, 
and therefore leaves the entropy unchanged. 
The quadratic part of the bare Hamiltonian is in general
\ba
  H = \frac{1}{2}\int\!\!d^3x \, 
  \left( Z_\pi \frac{\pi^2}{\kappa^2 \epsilon a^3} + 
  Z_{\zeta} \kappa^2 \epsilon a (\nabla \zeta)^2  \right) \, ,
\ea  
where $Z_{\pi}$ and $Z_{\zeta}$ renormalize the operators 
$\langle \pi(x) \pi(y) \rangle$ and $\langle \zeta(x) \zeta(y) \rangle$.
The commutation relations 
$\left[ \zeta(t,{\bf x}) ,\, \pi(t,{\bf y}) \right] = 
i \delta^{(3)}\left({\bf x} - {\bf y} \right)$ implies
\ba
  Z_{\pi} = \frac{1}{Z_{\zeta} }\, .
\ea
The renormalization of the wave function therefore leaves 
$\det(C)$ unchanged since the two products 
$\langle \zeta_{\bf k}(t) \zeta_{-{\bf k}}(t) \rangle \, \times \, 
 \langle \pi_{\bf k}(t)  \pi_{-{\bf k}}(t) \rangle$ and  
$\langle \left\{ \zeta_{\bf k}(t) ,\,  \pi_{-{\bf k}}(t) \right\} \rangle^2$
involve the same number of $\zeta$ and $\pi$.

The situation is a little different in a non renormalizable theory  
(still in Minkowski space) because the cancellation 
of divergences at the order $L$ of the loop expansion 
requires counterterms with higher 
derivatives. However at order $L$, these counterterms are local
and therefore do not change the entropy. 
In other words, the non-local 
part of the effective action (e.g. in Fourier space 
terms like $\ln(-k^2/\mu^2)$ or $\sqrt{-k^2/\mu^2}$) responsible
for the growth of entropy are precisely those
 predicted by the quantum theory \cite{Donoghue}. 
Irrelevant operators containing more that two derivative require some
care, since 
the equation of the two-point function is then higher 
than second order and can no 
longer be solved from the 
knowledge of the covariance matrix at an initial time.
These higher derivative terms are generally 
discarded in a self-consistent perturbative treatment, 
because the extra solutions of the 
propagation equation are not analytic in the limit $\hbar \to 0$ \cite{Simon}.
In this self-consistent perturbative sense, these counterterms do not
introduce ambiguities to the definition of the entropy.

Transposition to a curved background  
should not alter these conclusions
because the previous considerations about
the canonical structure, the distinction between 
local and non-local corrections and their distinct 
contributions to the entropy, and the role 
of higher derivative operators,
still hold on a curved background.
However, we have not verified explicitly 
for the models of paper II 
that the counterterms do not introduce ambiguities in a 
self-consistent perturbative treatment. We leave this analysis for future work.

\section{Separable density matrices as classical states}
\label{sec:sep}

To interprete the value of the entropy induced by 
a given coarse graining, we need a classification of the corresponding
reduced states. 
In the class of two-mode density matrices
there exists an operational definition of classical states.
After recalling the definition of separability, we 
explain why it cannot be applied unambiguously to cosmological perturbations
during the inflationary era.

\subsection{Definition and characterization of separable states}
\label{sec:defsep}

The state \Eqref{factorization} is entangled, i.e. it violates 
some Bell inequalities \cite{CPBell}.
We recall that this entanglement refers to the correlations between the 
modes of opposite wave-vectors. The definition of classicality 
called separability is a statement on the nature of these correlations.
Explicitely, separable states do not
violate {\it any} Bell inequality 
based on pairs of observables, each of them acting only in one sector, 
i.e. on ${\bf k}$ or on ${\bf -k}$. 
These states can therefore be represented as a convex sum of 
tensor products of density matrices  \cite{Werner},
\ba \label{defsep}
   \rho_{\rm sep} = 
   \sum_{l} p_l \, \rho_{{\bf k}}^{(l)} \otimes  \rho_{-{\bf k}}^{(l)} 
\, , \qquad p_l  \geq 0 \, .
\ea
Unlike entangled states, these can be prepared 
with local classical operations
(in the sense that one can prepare the 
states $\rho_{{\bf k}}^{(l)}$ and $\rho_{-{\bf k}}^{(l)}$ 
without making the modes ${\bf k}$ and $-{\bf k}$ interact), 
and by using a random generator 
(characterized by the probabilities $p_l$).
Although physically transparent, this definition 
\Eqref{defsep} is of little practical use because
of the difficulty of proving or disproving the existence of
the set of $\rho^{(l)}$ and $p_l$.
Fortunately, for Gaussian states, a 
criterion of separability is known \cite{Gsep}. 
For the GHDM, it is expressed as the following inequality on 
the parameter $\delta$ \cite{CPfat}
\ba \label{separability}  
  \rho_{\rm red} \,\,\, {\rm separable} 
  \quad &\Longleftrightarrow& \qquad \delta=1 \, .  
\ea
In brief, GHDM fall into two disjoint classes, 
separable states ($\delta \geq 1$) which are operationally indistinguishable
from stochastic ensembles,
and entangled states ($\delta < 1$) 
characterized by a departure of the anticommutator from classical correlations.
This is the definition of a classical GHDM  
we adopt. 

Finally, at the threshold of separability, 
the entropy is equal to one half of the 
entropy of the thermal state with the same value of $n$ (for $n \gg 1$),
see Eq. \Eqref{Sdelta}
\ba \label{entropydelta1}
   S(n,\delta = 1) = \frac{1}{2} S_{\rm Max}(n) \simeq \ln(n) \, . 
\ea

\subsection{Discussion}
\label{sec:sepdiscuss}

From \Eqref{n+1/2}, we see that the values of 
$n$ and $c$, and therefore that of $\delta$ as well, 
depend of the choice of canonical variables.
As explained in Section \ref{sec:intrinsic},
only the combination
\ba
  \det(C) = \left( n+\frac{1}{2}   \right)^2 - \vert c \vert^2\, ,
\ea 
does not. This poses a fundamental limitation to the applicability of 
the criterion of separability to systems with a time dependent Hamiltonian,
and therefore to the cosmological perturbations during inflation, 
see \Eqref{Hquad}.

For instance, in place of the variables 
$(\varphi_{\bf k}, \pi_{-\bf k} = \partial_{\eta} \varphi_{\bf k})$
and the Hamiltonian \Eqref{Hquad}, 
one could choose to work with the variables $\varphi_{\bf k}$ and 
$\tilde \pi_{-\bf k} = 
\partial_{\eta} \varphi_{\bf k} - (z'/z) \varphi_{\bf k}$, 
as done for instance in \cite{power}.  
In this case 
the corresponding Hamiltonian is 
\ba
H_{\bf k,\, -\bf k} = 
\vert \tilde \pi_{\bf k} \vert^2  + k^2 \vert \varphi_{\bf k}  \vert^2 + 
 \frac{\partial_{\eta}  z}{z} \left( \tilde \pi_{-\bf k} \varphi_{\bf k} 
+ \tilde \pi_{\bf k} \varphi_{-\bf k}\right)\, .
\ea
We leave it to the reader to verify that in the Bunch-Davis vacuum
the functional dependence of 
the variances is different from Eq. \Eqref{n+1/2} 
while $\det(C) = 1/4$. 
However, the canonical transformation 
from these two sets of variables 
mixes modes of opposite wave vectors,
i.e. it belongs to ${\rm Sp}(4)$ but not to its
subgroup ${\rm Sp}(2) \times {\rm Sp}(2)$. 
The property of (${\bf k},-{\bf k}$)-separability is therefore 
not stable under such canonical transformations. 

As a result, the 
threshold of separability \Eqref{entropydelta1} depends, 
during inflation, on the choice of canonical variables.
The inequality $\delta \geq 1$ is in fact an inequality on the 
eigenvalues of the so-called partial transpose of 
the density matrix \cite{Gsep}.
Namely, the lowest eigenvalue
should be larger than the variance in the vacuum state. Hence, for any 
choice of the creation and annihilation operators 
$(a_{(i)}, a_{(i)}^{\dagger})$, 
it yields $\delta_{(i)} \geq 1$. 
Therefore a separable state in 
$i$-representation can be entangled in other  
representations related to it by a 
Bogoliubov transformation mixing the modes ${\bf k}$ and $-{\bf k}$.

In this we find another illustration of the well known fact that 
when the frequency varies, one looses some of the 
usefull characterizations of quantum states.
 Remember that in time dependent backgrounds
 there is no intrinsic definition of the
occupation number, even though the expectation value of the
stress tensor stays well defined. Here, even though the entropy
is well defined, the criterion aofclassicality is not clearly defined
as long as the frequency significantly varies (to be more precise,
outside the validity domain of the WKB approximation.) 
 Since there is no
clear notion of particle as pair creation
(or mode amplification) proceeds, there is no 
clear distinction between quantum and classical states, 
as illustrated by the criterion of separability.

On the contrary, during the radiation dominated era, this ambiguity disappears
since the mode frequency of 
the Mukhanov-Sasaki variable is constant (because $z = {\rm cte}$)
\cite{MukhaPhysRep}.
Then, defining classical states as 
the separable ones presents several advantages.
First, since it relies on the  
possibility of violating Bell 
inequalities, it is an operational definition.
Second, the 
separation between quantum and classical states is sharp. 
This contradicts the common
belief that the quantum-to-classical transition is fuzzy.
Being sharp, ''the time of decoherence''
is also precisely defined.  
Starting from an entangled state, 
this transition occurs at the time when $\delta$ 
crosses $1$.

\section{Other criteria of classicality} 
\label{sec:othercriteria}

We show the equivalence between separability 
and three other classicality criteria.
The latter suffer from the same ambiguities during inflation 
as the one of separability. 
They must therefore be compared to each other 
in the same representation, e.g. 
that defined in Sec. \ref{sec:linperturb}. 
We conclude by discussing a class of inequivalent criteria
which do not suffer from the above ambiguity but 
which are based on another arbitrary choice.

\subsection{Criteria equivalent to separability}
\label{sec:equivcriteria}

\subsubsection{The broadness of the Wigner representation}
\label{WisQ2}

A first alternative criterion was introduced in \cite{KieferDiosiWigner} 
for one-mode systems, generalized in \cite{DoddHalliwell} to two-mode systems, 
and applied to cosmological perturbations in \cite{SPKnew}. 
It rests on the observation that in many respects, 
the Wigner representation 
behaves like a probability density over phase space, 
except for the fact that it 
can have a finer structure than $\hbar$, 
and in particular can take negative values in small regions.

We recall that the Wigner function 
can be defined by (for a system with canonical variables $(q,p)$)
\ba
  W_\rho(q,p) \equiv \int\!\!\frac{d\Delta}{2\pi} \, 
  e^{ip\Delta} \, \rho\left(q-\frac{\Delta}{2}, q+\frac{\Delta}{2}\right)\, .
\ea
For Gaussian states, $W_\rho$ is Gaussian and its covariance matrix is 
$C$ of Eq. \Eqref{covmatrix}.
Although it is positive everywhere,
we see from Eq. (\ref{eigenvalues}b) that when $\delta < 1$, 
the variance of the subfluctuant variable $\Pi$ is smaller 
than the variance in the vacuum.  
The criterion of classicality therefore consists to ask that the 
Wigner representation does
not contain features that are smaller than those of the Wigner function 
in the vacuum state (this definition includes the case of negative values).
For GHDM, we therefore expect that this is the case when
\ba
\langle \Pi \Pi^{\dagger}  \rangle \geq \frac{1}{2}\, ,
\ea i.e. $\delta \geq 1 +O(1/n)$,
which is equivalent to the criterion \Eqref{separability} for large $n$.

This requirement is made more precise by asking 
that the Wigner representation of
$\rho_{\rm red}$ is broad enough to be the 
Husimi (or $Q$-) representation of some normalizable density matrix $\rho'$.
Indeed, the Wigner representation of a state $\rho'$ is mapped 
onto its $Q$-representation by a convolution with a Gaussian function of 
covariance matrix $\frac{1}{2}$, i.e.
\ba \label{conv}
   Q_{\rho'}(q,p) = \int\!\!\frac{dq'dp'}{2\pi} e^{-(q-q')^2 - (p-p')^2}
   W_{\rho'}(q',p')\, ,
\ea 
which explains why the Husimi-representation 
$Q_{\rho'}$ of $\rho'$ is a broader function than its Wigner representation 
$W_{\rho'}$. Moreover,
the Husimi-representation $Q_{\rho'}$ of any density matrix is positive 
because $Q_{\rho'}(q,p)$ is the expectation value of $\rho'$ in  
the coherent state $\ket{(q+ip)/\sqrt{2}}$.
We show in Appendix \ref{app:WisQ} that 
the GHDM $\rho$ verifying this condition are
\ba
  \delta \geq \delta_Q = 1 + \frac{1}{4n} = \delta_{\rm sep} + \frac{1}{4n}\, ,
\ea
as anticipated from the heuristic argument in the previous paragraph.
It is larger than the condition of separability, but only 
slightly, and  
the two criteria are equivalent in the limit $n \gg 1$ relevant for 
cosmological perturbations.

\subsubsection{P-representability}
\label{Prepr}

A second alternative definition of classicality 
is the requirement that $\rho_{\rm red}$ admits a 
$P$-representation as defined by Glauber \cite{CPPeyresq,CPfat}.
It means that the states can be written as a 
statistical mixture of coherent states,
\ba \label{Prepr}
  \rho_{P-repr} \equiv \int\!\!\frac{d^2v}{\pi}\frac{d^2w}{\pi} \, 
  P(v,w) \ket{v,{\bf k}}\bra{v,{\bf k}} \otimes 
  \ket{w,-{\bf k}}\bra{w,-{\bf k}} 
  \, .
\ea
where $P(v,w)$ is a normalizable Gaussian distribution. 
A $P$-representable state is obviously separable. 
In general, the converse is not true but it turns out that for GHDM
$P$-representability and separability are equivalent. Hence
a GHDM is $P$-representable if, and only if
\ba
  \delta \geq 1 = \delta_{\rm sep}\, .
\ea
To avoid any misunderstanding, we emphasize  
that this condition does not mean 
that coherent states are ''the pointer states''.
 These have only been used as a resolution of the identity. 
 We shall return to this point in the next Section.

\subsubsection{Decay of off-diagonal matrix elements}

A third alternative condition of classical two-mode states is 
provided by the decay of interference terms of 
macroscopically distinct states,
sometimes referred as Schr\"{o}dinger cat states. 
Since this decay is asymptotic, 
this criterion is less precise then the other three
criteria.
However it is qualitatively equivalent to them in the following sense.
We refer to Appendix D in \cite{CPfat} for details. 
The off-diagonal matrix elements of the density matrix of the pure state
\Eqref{BDSP} in, say the basis of coherent states are correlated
over a range $\propto n$. 
Using the coherent states as representative of 
semi-classical configurations of the field at a given time,  
the squeezed state is a linear superposition of 
macroscopically distinct semi-classical configurations.
As for the entropy, see \Eqref{entropydelta1}, 
the correlation length between off-diagonal matrix elements
is very sensitive to the value of $\delta$ in the range $[0,\,1]$ where it
decreases monotonously from $O(n)$ to $O(1)$ as $\delta$ increases. 
For $\delta \geq 1$, the correlation length
depends very slowly on $\delta$ and stays $O(1)$. 
Hence the decay of the correlations length also distinguishes
classical Gaussian states as those with 
$\delta \geq 1$.

\subsubsection{Adding one quantum incoherently}

There is a simple physical interpretation to the criterion $\delta \geq 1$.
Such a density matrix is obtained from the pure state by adding incoherently
one {\it quantum} on average \cite{CPPeyresq}. 
One obtains
\ba
  n \mapsto n' = n + \frac{1}{2} \, , \qquad \vert c \vert \mapsto 
  \vert c' \vert = \vert c \vert =
  \sqrt{n(n+1)} \, .
\ea 
We have split evenly the contribution of the quantum between each mode in order to 
preserve statistical homogeneity.
One obtains that the corresponding value of $\delta$ defined by
$\vert c' \vert^2 = n'(n'+1 - \delta)$ is
\ba \label{deltafrompurealpha}
  \delta = 1 + \frac{1}{2n} +O\left( \frac{1}{n^2} \right)\, ,
\ea
in agreement with the other criteria when $n \gg 1$. 
As an interesting side remark, 
remembering Eq. \Eqref{entropydelta1}, 
the fact that the entropy gain associated with this 
addition of one quantum is large
clearly establishes the fragile character of the quantum entanglement.

\subsection{Inequivalent criteria}

Let us now discuss a criterion \cite{SPKentropy,SPKnew} 
which is not equivalent to the above four criteria. 
Instead of adding incoherently one quantum, one can consider 
the entropy gain associated with the loss of one {\it  bit} of information.
Then, whatever the system is, the change of entropy is 
$S = \ln(2)$.
Applying this criterion to the standard inflationary distribution, 
the value of the parameter $\delta$ is
\ba
 \delta_{\rm one\,  bit}
 \simeq \frac{0.4}{n}\, .
 \label{inequ}
\ea 
The statistical properties of the 
corresponding density matrix are essentially 
the same as those of the original pure state
\footnote{It was noticed 
in \cite{SPKentropy} that adding one quantum
yields to $S = S_{\rm max}/2$, but its physical 
interpretation was not mentionned. 
In this early work, the authors
do not however refer to criteria of section \ref{sec:equivcriteria}
and 
prefer instead the criterion $S = \ln(2)$, referring to \cite{expdecoh}
for an experimental situation where decoherence is effective for such values. 
This reference is misleading. Indeed, 
the experimental observation of decoherence reported there 
concerns a system of a two-level atom in a cavity interacting 
with the electromagnetic field in a coherent state (the environment).
One finds that the interferences are blurred when one photon is exchanged.
Since the Hilbert space of the atom has two dimensions, 
the exchange of one photon between the system and the environment
is equivalent to the exchange of one bit. In inflationary cosmology
since $n \simeq 10^{100}$, this correspondence is lost.}.
For instance, Bell inequalities are still violated and
 the correlation length between off-diagonal matrix elements is 
still $O(n)$.

\subsection{Summary}

In conclusion, we have found that the various criteria 
of decoherence fall into two distinct classes.
On the one hand,
classical states with respect to the statistics of the anticommutator
 $\hat {\cal O} = \{ \varphi_{\bf k}(t),\,  \varphi_{- \bf k}(t') \}$
are the separable states with $\delta \geq 1$.
This criterion is ambiguous during inflation because it rests on 
a choice of canonical variables $\varphi_{\bf k}, \pi_{\bf k}$. 
On the other hand, a lower bound on the entropy  
$S \geq \ln(N)$ is intrinsic as only the value of the entropy is involved.
However, we could not identify any operator(s) 
to which this criterion might refer to.
The value of $N$ is therefore 
not dictated by any physical property of the state. 
(It might be provided by the resolution
of observational data, but this would confirm that it does not 
characterize the state of the system.)

\section{Decoherence and statistical mixtures}

The two classes of criteria presented above seem a priori 
rather different as the first class is based on the quantum properties
of the system, whereas the second class uses only the value of the entropy.
In this Section, we show that these two criteria can be 
incorporated into a single treatment.

\subsection{A picture of decoherence}
\label{sec:picture}

We need first to express two-mode density matrices has a tensor product of two 
one-mode density matrices. Indeed, recall that 
entangled states cannot be represented
in the form \Eqref{Prepr}. Instead
all GHMD can be decomposed as  
\ba \label{12decomp}
   \rho_{\bf k,\, -\bf k}(\delta) = \rho_1(\delta ) \otimes \rho_2(\delta )\, ,
\ea
where $1$ and $2$ refer to a separation of the Hilbert space into  
two sectors defined by the variables 
\ba
  \varphi_{1,2} \equiv \frac{\varphi_{\bf k} \pm i \varphi_{-\bf k}}{\sqrt{2}}\, .
\ea 
Because of homogeneity, the matrices $\rho_1$ and $\rho_2$ 
are characterized by the same covariance matrix $C$, 
 which moreover coincides 
with that of $\rho$. 
Explicitely, one has
\ba \label{C1}
  C_1 = C_2 = 
   \left(   
      \begin{array}{cc}
  \langle \varphi_1^2 \rangle  &  \frac{1}{2}\langle \{ \varphi_1,\,  \pi_1 \} \rangle  
\\
  \frac{1}{2}\langle \{ \varphi_1,\,  \pi_1 \} \rangle   &   \langle \pi_1^2 \rangle
      \end{array} 
   \right) = 
   \left(   
      \begin{array}{ll}
          {\cal P}_{\varphi}  &  {\cal P}_{\varphi \pi}  \\
          {\cal P}_{\varphi \pi}  &   {\cal P}_{\pi}
      \end{array} 
   \right)= C \, .
\ea
In this way, the properties of the state of cosmological 
perturbations have been 
encoded into two fictitious one-mode systems. 
The entanglement ($\delta < 1$) between modes of opposite wave vectors
$\bf k$ and $-\bf k$ reflects into the existence of two
sub-fluctuant variables $\Pi_{1,2}$  
as in Eq. (\ref{eigenvalues}b).

The question we address concerns the use of
 minimal Gaussian states  $\ket{(v, \xi)}$ of 
$\varphi_{1,2}$, i.e. squeezed coherent states, to represent 
the states $\rho_{1,2}(\delta )$  
as statistical mixtures in the following sense 
 \ba \label{reprxi}
   \rho_{1}(\delta) = 
   \int\!\!\frac{d^2v}{\pi} \, 
   P_{\xi}(v; \delta)\, \ket{(v,\xi)} \bra{(v,\xi)} \, ,
 \ea
that is, by summing only over the complex displacement $v$. 
The basis of states we use 
therefore have a common orientation and elongation which is 
fixed by the squeezing parameter 
 \ba\label{xi2}
 \xi = r \, e^{2i\theta_c}\, . 
\ea
Using the $1\sigma$ contour in phase space,  
the state $\ket{(v, \xi)}$ draws 
an ellipse of unit area, centered around 
$\left( \bar \varphi_1 \propto {\rm Re}(v), \, 
\bar \pi_1  \propto {\rm Im}(v)\right)$,
with a long axis (the superfluctuant direction)
$\langle \phi \phi^{\dagger} \rangle \propto e^{2r}$ making and angle
$\theta_c$ w.r.t. the horizontal $\varphi_1$-axis.
The latter is chosen for simplicity along the superfluctuant mode of $\rho_1$
defined at Eq. (\ref{eigenvalues}a). 
Hence in this ''frame'',
$\theta_c$ is the relative angle between the big axis of
$\rho_1$ and the big axis of $\ket{(v, \xi)}$, 
see Eq. (\ref{thetac}).

As shown in Appendix \ref{proofRepr}, 
the states $\rho_{1,2}(\delta )$ can be represented as in \Eqref{reprxi} 
for any value of $\delta$. More interestingly,
the choice of the basis vectors, which is parameterized by $\xi$,
is more limited when 
$\delta < 1$ in that the range of allowed values of 
$\theta_c$ belongs to 
a bounded interval
which shrinks to zero as $\delta \to 0$.
That is to say, the pure state
($\delta = 0$) admits only one representation, itself. 
As $\delta$ increases from zero, the 
allowed range of $\theta_c$ 
increases but is necessarily strictly smaller than $\pi/2$.
One can say that 
$\rho_1$ ''polarizes'' the pavement \Eqref{reprxi} 
along its superfluctuant mode.
In addition, the range of $r$ also 
increases and the distribution $P_{\xi}(v)$ becomes broader, 
which means that a growing number of families of minimal states can be used 
to represent $\rho_1(\delta)$.
For any $\delta < 1$, one must have $r>0$, i.e. the states $\ket{(v, \xi)}$
are necessarily squeezed
(and as we saw they tend to align with the big axis of $\rho_1$).
When the threshold of separability is approached, i.e. $\delta \to 1^{-}$, 
the lower bound of $r$ decreases to zero. In this limit
one can represent $\rho_{1}(\delta \geq 1)$ with coherent states.
 Notice also that as $\delta$ increases, 
 the distribution ${\cal P}_{\xi}(v; \delta)$  
 becomes broader (in $v$)   
 and at the threshold of separability does not have 
 any structure finer that a unit cell of phase-space. 
 In this we recover what was observed in subsection \ref{WisQ2}.

The fact that $\rho_1$ polarizes phase space 
is easy to understand {\it a contrario}. Indeed, consider the   
 minimal states which are squeezed  
 in the direction perpendicular to that of $\rho_1$, 
 i.e. $\theta_c = \pi/2$.  
 The corresponding spread in the subfluctuant variable $\Pi$ 
 of \Eqref{rotatedvar} is large. 
 A statistical mixture of these states  
 is necessarily spreaded out in 
 the $\Pi$ direction, i.e. $\langle \Pi \Pi^{\dagger} \rangle > 1/2$, 
 that is $\delta > 1$, see \Eqref{eigenPi}, and therefore $S > S_{\rm Max}/2$.

Let us formulate these results the other way around 
in terms of $\delta_\xi$ (or $S_{\xi}$), 
the amount of decoherence (or entropy)
needed for the state to be written 
in the $\xi$-basis as in Eq. (\ref{reprxi}).
According to what we just said,  
the further $\ket{(v,\xi)}$
departs from $\rho_1$, the larger is $\delta_{\xi}$
(and $S_{\xi}$). 
For well aligned $\ket{(v,\xi)}$, i.e. for $\theta_c = 0$,
$\forall \, \abs{\xi}=r \neq 0$, we have
$\delta_{\xi} < \delta_{\rm sep}$, the 
minimal amount of decoherence to reach the separability threshold.

Hence given a certain decoherence rate, these considerations 
translate into the time of decoherence $t_{\xi}$, 
i.e. the time $t_{\xi}$ after which {\it any} initial state $\rho_1(t_0)$ 
has evolved, as decoherence proceeds (as $\delta$ increases), 
into the statistical mixture \Eqref{reprxi} for that value of $\xi$.  
When using again well aligned $\ket{(v,\xi)}$, this 
defines $t_{\rm sep}$ as the maximal $t_{\xi}$. It is also
also the first time such that the representation \Eqref{Prepr} is allowed.

\subsection{Ambiguity in choosing pointer states}
\label{sec:pointer}

Pointer states are meant to bridge the gap between the quantum and classical 
descriptions of a system \cite{Kieferbook,Zurekreview}.
The defining property of pointer states is their robustness 
over a given lapse of time $t_{\rm p.s.}$ which makes them
the quantum counterparts of points in the phase space of classical mechanics.
That is, once the system has been prepared into a given initial state and
placed into contact with a given environment, they
are the states the least perturbed by the environement over 
$t_{\rm p.s.}$.
One generally obtains radically different pointer states 
whether $t_{\rm p.s.}$ is much smaller or comparable to the dynamical time 
scales of the open dynamics (the proper frequency of the system or
the characteristic time of dissipation).

To make the choice of $t_{\rm p.s.}$ 
less arbitrary, one often adds the requirement 
that any initial state of the system evolves, over a time $t_{D}$, 
into a density matrix which cannot be operationally distinguished from a
statistical mixture of the pointer states.
Since the evolution of both the pointer states and $\rho$ 
are governed by the same dynamics, consistency requires that 
the times $t_{\rm p.s.}$ and $t_D$ be commensurable.

As we saw at the end of Sec. \ref{sec:picture},
choosing a family of pointer states (i.e. a pair $(r,\theta_c)$),
is arbitrary since it amounts to a choice of $\delta_\xi$, or
to a choice of a lower bound $S \geq \ln(N_\xi)$
which we recall is not dictated by physical considerations.
The corresponding time of decoherence $t_D = t_\xi$ is therefore arbitrary. 
But there is a more fundamental obstruction to the identification 
of a pointer basis
during inflation, namely pointer states refer to a choice of canonical 
variables, which is arbitrary when the Hamiltonian depends 
explicitly on time (compare for instance narrow wave-packets in 
$\varphi$ or $\zeta$).

In brief, on the one hand 
during inflation, the question of finding "the" pointer states
is not well-defined. 
On the other hand, 
during the radiation dominated era,
the criterium of separability offers an unambiguous definition 
of classical states based on the statistical properties of the state.
From these two facts we conclude that the concept of 
pointer states does not seem useful to analyse the decoherence
of cosmological perturbations.

\section{Summary}

By truncating the hierarchy of Green functions, 
we first show how to get 
a reduced density matrix for the adiabatic perturbations, 
in a self-consistent manner, and from the interacting 
properties of the system itself, i.e. without introducing 
some {\it ad hoc} environmental degrees of freedom.

When truncating the hierarchy at the first non trivial level,
statistical homogeneity
still implies that the density matrix factorizes into
sectors of opposite wave vectors.  
Hence the reduced density matrix of each sector is 
determined by three moments
related to the anti-commutator function, see \Eqref{covmatdressed}.
This also implies that decoherence
here describes the loss of the entanglement of these two modes.
The level of decoherence is characterized by one 
real parameter in each sector, the parameter  $\delta$ introduced in 
\Eqref{ldef}.

We then show that the entropy $S$ contained in each 
reduced two-mode density matrix is a well-defined quantity 
which monotonously grows with $\delta$. 
The important conclusion is that
$S$ is the only intrinsic quantity of these reduced density matrices, 
in that the other quantities all require to have chosen some pair 
of canonical variables to be evaluated. 

After the entropy, we studied the quantum-to-classical transition.
We show that the criterion of separability agrees with three other
criteria, namely the broadness of the Wigner function, 
the $P$-representability,
and the neglect of off-diagonal elements of the density matrix. 
During inflation, these four concepts are ill defined, 
because, contrary to the entropy,
they require to have selected some creation and destruction operators, 
which is ambiguous since the frequency significantly 
varies when modes are amplified.

We compare the above four criteria which give a large entropy 
at the threshold of classicality, see \Eqref{entropydelta1},
to another class of criteria which give much smaller entropies,
see \Eqref{inequ}, and explain why they differ so much. 
In the last Section, we present a unified treatment of the reduced
density matrices in which both types of criteria can be found.
We show that each reduced density matrix can be written as 
statistical mixtures of minimal states which posses a well 
defined range of the squeezing parameter $\xi$ of Eq. \Eqref{xi2}. 
The details are given in Appendix \ref{proofRepr}. 

This analysis clearly shows that there is no intrinsic 
definition of a threshold of decoherence  
(or a critical entropy)
at which the quantum-to-classical transition would occur during inflation.
We also argue that the pointer states are of no help in providing
such a definition.

In the next paper, 
we calculate the entropy for two dynamical models.
It appears that no significant entropy 
is gained during single field inflation,
so that the quantum-to-classical transition should occur 
during the adiabatic era. 
On the contrary, it is very efficient in multifield scenarios.

\acknowledgments{The work of D.C. is supported by the Alfried Krupp Prize for Young 
University Teachers of the Alfried Krupp von Bohlen und Halbach
Foundation.}

\begin{appendix}

 \section{Why bother about the covariance matrix ?}
 \label{sec:whybother}

 In this Appendix, we emphasize the role of the dynamics and the
 canonical structure in defining reduced states and their associated entropy.
  
 One may object to our  
 analysis of the two-point function at different  
 times on the grounds that 
 one can only measure the values of this correlation function
 on our past light cone or on the last scattering surface.
 This objection can be answered as follows. The outcomes of a measurement 
 are analyzed through the grid of a particular dynamical model. 
 There are in general more
 dynamical variables necessary to write a consistent model than one can 
 actually measure. 
 The state of the system, either the exact state or 
 a partial reconstruction of it, 
 depends on both the outcome of the measurement 
 (for instance, the values of the 
 power spectrum and bispectrum) and the correlation
 functions.

 As an example, we keep only the power spectrum and show that
this assumption is untenable
 as it amounts to ignore the canonical structure of the theory.
 As a result the von Neumann entropy of the reconstructed 
 state is not well defined.
 Indeed, following the algorithm of Sec. \ref{sec:opdef}, 
 we write the {\it Ansatz} for the density matrix 
 \ba \label{ansatz2}
   \rho_\zeta &=& \frac{1}{Z_\zeta} \exp\left( - 
   \lambda \frac{k^3}{2\pi^2} \hat \zeta^2 \right)
   = \int\!\!d\zeta \, P(\zeta) \ket{\zeta}\bra{\zeta}
   \, , \qquad 
   Z_{\zeta} = \sqrt{\frac{\pi}{\lambda}} \, ,
 \ea 
 where $\lambda$ is a Lagrange multiplier
 ensuring that the power spectrum of $\zeta$ has the measured value 
 $P_{\zeta}$, i.e. ${\cal P}_{\zeta}(q) = 
 {\rm Tr} \left(\rho_{\zeta} \hat \zeta_{\bf q} \hat \zeta_{-\bf q} \right)
  = 1/2\lambda$. The distribution $P(\zeta)$ is therefore
 \ba \label{Pansatz2}
   P(\zeta) = \frac{1}{\sqrt{2\pi {\cal P}_{\zeta}}} 
   \exp\left( -\frac{k^3}{2\pi^2}\frac{\zeta^2}{2 {\cal P}_{\zeta}}   \right)\, .
 \ea
  and the von Neumann entropy is then found to be
 \ba \label{stupidentropy}
     S = \ln Z_\zeta  + \lambda {\cal P}_{\zeta} 
     = \frac{1}{2}\ln\left( 2\pi {\cal P}_{\zeta} \right) + \frac{1}{2} \, .
  \ea
 The constant $1/2$ is universal and corresponds in this scheme to the 
 entropy of 
 the vacuum.  
 For ${\cal P}_{\zeta} \sim 10^{-10}$, 
 the entropy \Eqref{stupidentropy} is negative.
 To cure these pathologies, we must add some physical input, i.e. enlarge the 
 set of observables
to include the variances ${\cal P}_{\pi}$ and ${\cal P}_{\zeta \pi}$.

Another way to see the necessity to consider $C$ rather than 
${\cal P}_{\zeta}$ alone is the following.
 Note that even though the state \Eqref{ansatz2} is Gaussian, 
 we cannot calculate its von Neumann entropy from the formulas
 \Eqref{covmatrix}, \Eqref{SvN} and \Eqref{SvN2}.
 The latter are only valid for a state reconstructed 
 from the anticommutator function.
 If one insists on doing so, one finds ${\cal P}_{\pi} = \infty$ (since
 $\rho_\zeta$ is diagonal in the field-amplitude basis), so that
 the entropy is infinite.
 The state $\rho_\zeta$ must be therefore be regularized first, 
 by giving a finite 
 width to $\langle \pi^2 \rangle$. This operation is arbitrary 
 without additional physical input about the dynamics.

\section{Equivalence of the criteria of separability and 
of broadness of the Wigner function}
\label{app:WisQ}

Let us consider a Gaussian density matrix $\rho$ of a bipartite system
and let $C$ be its covariance matrix.
We adopt a different parameterization as in the text, following 
\cite{Gsep}
\ba
  C = {\rm Tr}\left( \rho \,  \left\{ A,\, A^{\dagger} \right\}  \right)
    = \left(  
   \begin{array}{cccc}
          n+\frac{1}{2} & 0 & 0 & c \\
          0 & n + \frac{1}{2} & c^* & 0 \\
          0 & c & n+\frac{1}{2} & 0  \\
          c^* & 0  &  0 & n + \frac{1}{2}  
   \end{array} 
   \right)\, ,
 \qquad 
  A =     
   \left(
   \begin{array}{l}
           a_{\bf k} \\
           a_{\bf k}^{\dagger} \\
           a_{-{\bf k}} \\
           a_{-{\bf k}}^{\dagger}  
   \end{array} 
   \right)\, .
\ea
The positivity of the density matrix and the non-commutativity of the
creation and annihilation operators
puts a constraint on $C$, namely
\ba \label{positivity}
 C + \frac{E}{2} \geq 0 \, , \qquad 
 E =   
   \left(
   \begin{array}{cccc}
           1 & 0 & 0 & 0 \\
           0 & -1 & 0 & 0  \\
           0 & 0 & 1 & 0  \\
           0 & 0 & 0 & -1  
   \end{array} 
   \right)\, .
\ea
This is the general form of the Heisenberg uncertainty relation for the second
moments.
Indeed, this condition puts a lower bound on the lowest 
eigenvalue of $C$, which, in the case of homogeneous states, reads 
$n+1/2 - \sqrt{\abs{c}^2 + 1/4} \geq 0$. The latter can be
recast as the Heisenberg uncertainty relation
\ba
  \langle a_{\bf k} a_{\bf k}^{\dagger} \rangle  
  \langle a_{\bf k}^{\dagger} a_{\bf k} \rangle  \geq 
  \abs{\langle a_{\bf k} a_{\bf k} \rangle }^2\, .
\ea

A necessary and sufficient condition for the separability of 
a GHDM \footnote{We recall that a necessary 
condition for $\rho$ to be a separable density matrix is 
that its partial transpose is a {\it bona fide} density matrix \cite{Peres}.
The partial transpose 
$\rho_{\rm pt}$ is by definition, obtained from $\rho$ by 
a transposition in one sector only, say 
$-{\bf k}$. In any basis $\ket{n, {\bf k}}\ket{m,\, -{\bf k}}$, this is
 written as
$\bra{n, {\bf k}}\bra{m,\, -{\bf k}} \, 
  \rho_{\rm pt} 
 \ket{m',\, -{\bf k}} \ket{n', {\bf k}} = 
 \bra{n, {\bf k}}\bra{m',\, -{\bf k}} 
  \rho 
 \ket{m, -{\bf k}}\ket{n',\, {\bf k}}$. 
In Eq. \Eqref{CNseparability} this operation is implemented 
by the matrix $\Lambda$.
Gaussian states are separable when this condition is satisfied \cite{Gsep}.} 
is 
\ba \label{CNseparability}
  \rho \,\, {\rm separable} 
 \qquad \Longleftrightarrow \qquad 
   \Lambda C  \Lambda + \frac{E}{2} \geq 0  \, ,
  \qquad \Lambda =   
   \left(
   \begin{array}{cccc}
           1 & 0 & 0 & 0 \\
           0 & 1 & 0 & 0  \\
           0 & 0 & 0 & 1  \\
           0 & 0 & 1 & 0  
   \end{array} 
   \right)\, ,
\ea
which gives $n - \abs{c} \geq 0$, or $\delta \geq 1$ using (\ref{ldef}).

A different criterion than separability is used in
\cite{KieferDiosiWigner}. 
With this condition, the 
Wigner representation of $\rho$ must be broad enough to be also
the $Q$-representation of a density matrix $\rho'$.
We specialize to Gaussian states.
To obtain the corresponding condition on the Wigner
function, one simply notices that 
the convolution of the Wigner representation of a state $\rho$ 
with a Gaussian of covariance matrix $1/2$ 
yields the $Q$-representation of that state, 
\ba \label{convolution}
  &&W_{\rho}(V) =  \frac{1}{\sqrt{{\rm det} C}} \, 
  \exp\left( -\frac{1}{2} V^{\dagger} C^{-1} V \right) 
\, \mapsto \,
Q_{\rho}(V) = \frac{1}{\sqrt{{\rm det} C_Q}}\, 
  \exp\left( -\frac{1}{2} V^{\dagger} C_Q^{-1} V \right) \, ,
\nonumber \\  
  && \qquad   \qquad 
  {\rm with }\qquad   \qquad 
  C \quad \mapsto\quad C_Q = C + \frac{1}{2} \, ,
\ea
where 
$V^{\dagger} = \left( \varphi, \pi^\dagger, \varphi^\dagger, \pi  \right)$.
Hence the condition \Eqref{positivity} on $C$
can be written as a similar condition for the covariances $C_Q$ 
of the $Q$-representation
\ba
  C + \frac{1}{2}\left(E - 1  \right) \geq 0 \, .
\ea
In consequence, a necessary condition for the Wigner 
representation $W_{\rho}$ of a state $\rho$ to be also the 
$Q$-representation $Q_{\rho'}$ of a state $\rho'$ is that the 
covariance matrix $C$ of $\rho$ verifies 
$C + \frac{1}{2}\left(E - 1 \right) \geq 0$.
For Gaussian states, this condition is also sufficient. 
Specializing now to the case of GHDM, we arrive at the conclusion
\ba \label{WisQ}
 n^2 - 1/4 \geq \abs{c}^2 
\quad \Longleftrightarrow \quad 
\delta \geq 1 + \frac{1}{4n}\, . 
\ea
This is slightly more 
constraining than the separability condition $n \geq \abs{c}$, 
but the difference between the two 
is not relevant when $n \gg 1$.

\section{Representations of partially decohered distributions}
\label{proofRepr}

In this Appendix we show that all partially decohered Gaussian distributions
can be written as statistical mixtures of minimal states which belong
to a certain family.
As decoherence increases, the ranges of the parameters 
characterizing this family become larger. As the threshold
of separability, the angle between the super-fluctuant modes
of the minimal states and that of the distribution becomes unrestricted.

The statistical homogeneity allows a formal reduction of the problem.
One can decompose the field amplitude $\phi_{\bf q}$ into its 
''real'' and ''imaginary'' parts 
$\phi_{\bf q} = (\phi_1 + i \phi_2)/\sqrt{2}$, 
such that
\ba
   a_{1} = \frac{1}{\sqrt{2}} \left( a_{\bf k} + a_{-\bf k}  \right) \, , \qquad 
   a_{2} = \frac{-i}{\sqrt{2}} \left( a_{\bf k} - a_{-\bf k}  \right)\, .
\ea
With this decomposition of the Hilbert space, GHD factorize
\ba
   \rho_{\bf k,\, -\bf k} = \rho_{1} \otimes \rho_{2} \, ,
\ea
where $\rho_{1} = \rho_{2} $. In addition, 
the parameters of \Eqref{defn}  
are given by
\ba \label{netc}
   n  
   = {\rm Tr}\left(\rho_1\,  a_{1}^{\dagger} a_{1} \right) \, , \qquad 
   c 
   = {\rm Tr}\left(\rho_1\,  a_{1}^2 \right)\, .
\ea 
Similarly, for two-mode coherent states 
$\ket{v, \bf k} \otimes \ket{w,-\bf k} = \ket{v_1} \otimes \ket{v_2}$ 
where $v=(v_1 + i v_2)/\sqrt{2}$ and 
$w=(v_1 - i v_2)/\sqrt{2}$.

Let us consider a Gaussian density matrix $\rho$ 
 of a single mode, characterized by a value of $0 \leq \delta \leq n+1$.
 We ask whether there exists a family of minimal Gaussian states 
$\ket{(v,\xi)}$
and a Gaussian distribution $P_{\xi}(v)$     
such that
 \ba \label{repr1mode}
   \hat \rho_{\delta} = \int\!\!\frac{d^2v}{\pi}\,
   P_{\xi}(v)\, \ket{(v,\xi)}\bra{(v,\xi)} \, .
 \ea
The minimal Gaussian states are chosen to be the displaced squeezed states 
\ba
\ket{(v,\xi)} = D(v) S(\xi) \ket{0}\, ,
\ea
 where $D(v) = e^{(va^{\dagger} - v^* a)}$ is the
displacement operator and 
\ba
S(\xi) = e^{\frac{r}{2}(e^{i\theta}a^{\dagger \, 2} - e^{-i\theta} a^2)}\, ,
\ea 
the squeezing operator. 
The distribution $P_{\xi}(v)$ is centered and is therefore
defined by its covariance matrix $C_\xi$ 
\ba
   P_{\xi}(v) &=& \left({\rm det} C_\xi\right)^{-1/2}   
   \exp\left\{ - \frac{1}{2} X^\dagger C_\xi^{-1} X \right\}  ,
   \qquad 
   X = 
    \left(
   \begin{array}{l}
           v \\
           v^* 
   \end{array} 
   \right) ,
 \qquad
   C_{\xi} =  
    \left(   
      \begin{array}{ll}
          n_\xi      & c_{\xi}\\
          c_{\xi}^*  &  n_{\xi} 
      \end{array} 
   \right) .
\qquad 
\ea 
The moments of the state $\rho$ defined at Eq. \Eqref{defn}
are obtained from 
the ones of the distribution $P_{\xi}$
by the definition \Eqref{netc}
\ba
   n &=& \int\!\!\frac{d^2v}{\pi}\,
   P_{\xi}(v)\, \bra{(v,\xi)}\, a^{\dagger} a \, \ket{(v,\xi)} = 
   \int\!\!\frac{d^2v}{\pi}\,
   P_{\xi}(v)\, \left( \abs{v}^2 + \abs{\beta}^2 \right)\, ,
 \nonumber \\
   c &=& \int\!\!\frac{d^2v}{\pi}\,
   P_{\xi}(v)\, \bra{(v,\xi)}\, a^2 \, \ket{(v,\xi)} = 
   \int\!\!\frac{d^2v}{\pi}\,
   P_{\xi}(v)\, \left( v^2 + \alpha\beta \right)\, .
\ea
Here, $\alpha$ and $\beta$ are the Bogoliubov coefficients 
associated with the squeezed state $\ket{\xi}$ by 
\ba
  \alpha = \ch(r)  \, , \qquad \beta = e^{-i2\theta_c} \sh(r)
  \, , \qquad \xi = r e^{i\theta_c}\, .
\ea
Hence, the momenta of the distribution $P_{\xi}$ are
\ba
  n = n_\xi + \abs{\beta}^2  \, ,\qquad 
  c = c_\xi + \alpha\beta \, .
\ea
The only constraints on $n_{\xi}$ and $c_{\xi}$  
arise from 
the fact that the right hand side of \Eqref{repr1mode} must be positive and
normalizable in order to be a density matrix.
These conditions are respectively
\ba \label{constraindet}
  \rho \geq 0 \qquad &\Longleftrightarrow& \qquad P_{\xi} \geq 0 \, ,
  \nonumber \\
  {\rm Tr}(\rho) = 1 \qquad &\Longleftrightarrow& \qquad
  \int\!\!\frac{d^2 v}{\pi} \, P_{\xi}(v) = 1 
  \qquad \Longleftrightarrow \qquad 
  {\rm det}(C_{\xi}) \geq 0
\ea  
We write 
\ba
  \xi=re^{i\left(2\theta_c + \arg(c)\right)}\, ,
  \label{thetac}
\ea 
where $\theta_c$ 
is the angle made by the 
squeezed coherent states with the super-fluctuant mode (the eigenvector of 
$C$ with the largest eigenvalue). 
We now look for the range 
of values of $x = e^{2r}$ and $\theta_c$ allowed by the 
constraint on the determinant \Eqref{constraindet}, that is
\sba \label{inegalityx}
 R(x) &=& Ax^2 + 2Bx + C \leq 0 \, , \\
  A &=& n + \frac{1}{2} - \abs{c} \cos(2\theta_c) \, ,
 \quad 
  B = - \left(n\delta + \frac{1}{2} \right) \, ,
\quad  
  C = n + \frac{1}{2} + \abs{c} \cos(2\theta_c) \, .
\qquad 
\sea
Since $A > 0$, the inequality (\ref{inegalityx}a) 
can only be satisfied if the discriminant
$\zeta = B^2 - AC$ is positive. 
Since the coefficients 
$A$, $B$ and $C$ depend only on $\theta_c$ 
and $\delta$, this gives an implicit constrain equation for 
$\theta_c(\delta)$,
\ba 
  \zeta(\theta_c, \delta) = (n\delta)^2 - \abs{c}^2 \sin^2(2\theta_c) \geq 0
  \qquad \Longleftrightarrow \qquad 
  \sin^2(2\theta_c) \leq 
  g(\delta) \equiv 
  \frac{(n\delta)^2 }{n(n+1 - \delta)} \, .
\qquad 
\ea
The function $g(\delta)$ is strictly growing over the interval 
$0 \leq \delta \leq n+1$ and takes the special values 
$g(0)=0$ and $g(1) = 1$. 
We distinguish the two following cases: \\ 
1) If $\delta \geq 1$ (the two-mode state from which $\rho$ is obtained is 
separable), all the values of $\theta_c$ are allowed and the squeezed 
coherent states in 
\Eqref{repr1mode} can have an arbitrary orientation.
\\
2) If $0 \leq \delta < 1$, the angle $\theta$ can only vary into the interval
$\left[ -\theta_M , + \theta_M \right]$ where the angular opening is defined by
\ba
  \sin^2(2\theta_M)  \equiv  \frac{(n\delta)^2 }{n(n+1 - \delta)} \, .
\ea
The value $\theta_M = \pi /2$ corresponds to $\delta = 1$.
The interval $\left[ -\theta_M , + \theta_M \right]$ shrinks to zero 
as $\delta$ decreases.
In the limit $\delta = 0$ (that is, for the pure squeezed state), the
squeezed coherent states $\ket{(v,\xi)}$ must be aligned 
with the superfluctuant mode of $\rho_1$.

We now specialize to the case 
$0\leq \delta \leq 1$ and we look for the range of allowed values of 
$x$ for a given $\delta$ and $\theta$. 
We first note that 
\ba 
  1 \leq x \leq x_M \equiv 2\left[ \, n+\frac{1}{2} + \sqrt{n(n+1)} \right]\, .
\ea 
The lower bound 
comes from $r = \abs{\xi} \geq 0$, and the upper bound 
from the requirement that 
the eigenvalues of $C_{\xi}$ (which are expectation values) 
are positive, that is 
$n_{\xi} \geq \vert \beta \vert^{2} =\sh^2(r)$. 
For the values of $\theta_c$ such that the discriminant $\zeta \geq 0$, 
 $R(x)$ of Eq. (\ref{inegalityx}a) 
 is negative for $x$ in the interval $[x_-,\, x_+]$, where the roots of the 
polynome are 
\ba
  x_{\pm}(\theta, \delta) &=& \frac{1}{A(\theta)} 
  \left\{ \frac{1}{2} + n\delta \pm  
  \left[ (n\delta)^2 - \abs{c}^2 \sin^2(2\theta) \right]^{1/2} \right\}\, .
\ea
For a given value of $\delta$, the length of the interval is a
strictly decreasing function of 
$\theta_c$: the smaller the deviation between the super-fluctuant 
directions of 
$\ket{(v,\, \xi)}$ and $\rho$, 
the smaller the allowed range of $x$. 
Indeed, the smallest root decreases from
$x_{-}(0;\delta) = 2[n+1/2 + \abs{c}(\delta)]/(1+ 4n\delta) \geq x_M$ to 
$x_{-}(\theta_M;\delta) = (1+2n\delta)/2A(\theta_M)$, 
while the largest root decreases from
$x_{+}(0;\delta) = 4[n+1/2 + \abs{c}(\delta)] > x_M$ 
to 
$x_{+}(\theta_M;\delta_R) = x_{-}(\theta_M;\delta_R) < x_M$.

The common value of the roots at $\theta_M$ means that for this angle, 
there is a 
unique value of the squeezing parameter. Not surprisingly for $\delta = 1$, 
this common value is $1$, i.e. $r =0$, and
 the fact that $\theta$ can take any value corresponds to the isotropy of the 
coherent states.

To study the limit $\delta \to 0$, please notice that we can set 
$\theta = 0$ and that $x_+(0,\delta)$ is larger that
$x_M$ for $\delta \ll 1$.
It means that $x$ bellongs to the interval $[x_-(\theta), x_M]$ 
which shrinks to $x_M$ since 
$x_{-}(0;\delta)$ approaches $x_M$ from below.
In the limit of a pure state ($\delta = 0$),
there is a unique representation of the form \Eqref{repr1mode},
the state itself (that is the distribution of Eq. (\ref{repr1mode}) is
$P_{\xi} = \delta_{\rm Dirac}^{(2)}\left( v_1  \right)$).

\end{appendix}

\end{document}